\documentstyle[prl,aps,twocolumn]{revtex}

\draft

\author{M. A. Nielsen \thanks{Electronic address :
mnielsen@tangelo.phys.unm.edu}}
\title{Chaos in the Quantum Measurement Record \thanks{Submitted to Physical
Review Letters.}}
\address{Department of Physics, University of Queensland 4072, Australia
\thanks{New address : Center for Advanced Studies, Department of Physics and
Astronomy, University of New Mexico, Albuquerque NM 87131-1156}}

\date{\today}

\begin{document}

\maketitle

\begin{abstract}
We investigate measures of chaos in the measurement record of a
quantum system which is being observed. Such
measures are attractive because they can be directly connected to experiment.
Two measures of chaos in the measurement record are defined and
investigated numerically for the case of a quantum kicked top. A smooth
transition between chaotic and regular behavior is found.
\end{abstract}



Understanding how chaotic behavior arises in quantum systems
has been the subject of much research in recent years
\cite{Gutzwiller90a,Reichl92a}. The purpose of this Letter is to
investigate measures of chaos which are
determined wholly by the {\em measurement record} of a quantum
system which is undergoing observation. Such measures
are attractive because they can be directly connected to experiment.
It is well known that the dynamics of a quantum system
depend on how that system is observed. This is in contrast to
classical systems, for which it is usually assumed that measurements
can be performed without disturbing the system. Thus we are led to
a view of chaos in which the initial state, the
free (Hamiltonian) dynamics, and the observation scheme are all
regarded as parameters which can be varied in order to gain a better
understanding of chaos in specific systems. 

The Letter is organized as follows. We begin by reviewing some ideas from
algorithmic information theory, which are then used to define a measure
of chaos in the measurement record. A second measure of chaos in the
measurement record is defined using the Shannon entropy. An inequality
between these two measures is proved. A numerical investigation of
these measures is then performed for the specific example of the
quantum kicked top. We conclude with some general observations and
comments on how the present approach to quantum chaos may be
extended.

Suppose a series of measurements is made on a system, either classical
or quantum. The result is a sequence of
results, $R_1,R_2,\dots$, which we call the {\it measurement record}.
An intuitively appealing measure of chaos is the
rate of growth of the information needed to describe the measurement record.
If this occurs very quickly, then the measurement
record is difficult to describe, and we would say that the system is
chaotic. If the rate of information growth is
slow, then the measurement record is easy to describe, and
we would say that the system is regular.

To make this idea precise, we need to know how to quantify the
``information needed to describe the measurement record''.
The solution to this problem is known as
{\em algorithmic information theory}, which was created independently by
Solomonoff, Kolmogorov and Chaitin
\cite{Solomonoff64a,Kolmogorov65a,Chaitin92a}. They defined
the {\em algorithmic information content} ${\cal I}(s)$ of
a string $s$ to be the length of the shortest possible description of that
string. More rigorously, ${\cal I}(s)$ is the length of the
shortest {\it algorithm} which generates the string $s$. An algorithm,
as defined by Turing \cite{Turing37a}, can be represented as a {\it program}
for a {\it universal computer}, $U.$ A data string $s$ is used as input to such
a program $p$, which then performs the algorithm on the data string.
If the program ever halts then it prints an output string, $U(p,s)$.

For a particular universal computer $U$ we define the
{\em algorithmic information content} ${\cal I}_U(s)$ of a string $s$ to be the
minimum length $|p|$ taken over all programs $p$ for $U$ such that
$U(p,\emptyset) = s$, that is the program $p$ prints out $s$ when
nothing is input. The {\em conditional algorithmic information content} 
${\cal I}_U(s|t)$ of $s$ given another string $t$ is defined to be the minimum
length $|p|$ taken over all programs $p$ such that $U(p,t) = s$. Note that
here and elsewhere it is assumed for convenience that $p$ and $s$ are
represented as binary strings. In
chapter six of \cite{Chaitin87a} it is proved that
\begin{eqnarray} \label{aic inequality}
{\cal I}_U(s|t) \leq {\cal I}_U(s) + o(1), \end{eqnarray}
where $o(1)$ is an order one constant which does not depend on $s$ or $t$.
This intuitively reasonable inequality is important in our later reasoning.

We now turn briefly to classical systems, using algorithmic information
theory to define a measure of chaos, and seeing how this measure relates
to other measures of chaos for classical systems.
Suppose a classical system is initially at a point $x_0$ in some phase
space. At times $t_1,t_2,\dots$ a measurement is carried out on the system.
We suppose that this measurement can be modelled as follows : There is
a partition of cells covering phase space, labelled $1,\dots,n$ and the
measurement at time $t_i$ determines which of these cells, $R_i$, the system is
in at that time. Define the {\em rate of information production}
of the classical system to be
\begin{eqnarray}
{\cal R}(x_0) := \limsup_{n \rightarrow \infty}
   \frac{{\cal I}(R_1,...,R_n)}{n}. \end{eqnarray}
A minor technical point is that this definition is independent of which
universal computer $U$ is used to define algorithmic information,
and thus no subscript $U$ is needed on ${\cal I}$ or ${\cal R}$.
This follows from the fact \cite{Chaitin87a} that for any two
universal computers $U_1$ and $U_2$, there is a constant $C$ such that
for any string $s$,
\begin{eqnarray}
|{\cal I}_{U_1}(s) - {\cal I}_{U_2}(s)| \leq C. \end{eqnarray}

It can be proved (see \cite{Alekseev81a} for a review and references) that
\begin{equation} {\cal R}(x_0) \leq h, \end{equation}
where $h$ is the topological entropy of the system. Positivity of
the topological entropy is widely regarded as one of the best
indicators of chaos in a classical system, and thus we can see that
if ${\cal R}(x_0)$ is ever positive then the topological entropy
is positive and thus the system is chaotic.

Consider now the case of a quantum mechanical system which is 
observed at times $t_1,t_2,\dots$, producing a measurement record
$R_1,R_2,\dots$. Again, the rate of information production may be defined
as
\begin{eqnarray}
{\cal R} := \limsup_{n \rightarrow \infty} \frac{{\cal I}(R_1,...,R_n)}{n}.
\end{eqnarray}
In general the sequence $R_1,R_2,\dots$ is a random sequence because
of the inherently probabilistic nature of quantum measurements.
In order to estimate the rate of information production we use the result
(see \cite{Schack94b} for recent results and references) that
the average algorithmic information of a random sequence
$R_1,\dots,R_n$ given background information
$B = \{ p_{r_1,\dots,r_n} \}$
consisting of the probabilities of each possible measurement record
$r_1,\dots,r_n$ satisfies the inequality
\begin{eqnarray} \label{aic approx}
{\cal H}_n \leq \overline {\cal I}(R_1,\dots,R_n|B) \leq {\cal H}_n + o(1), 
\end{eqnarray}
where ${\cal H}_n$ is a form of the Shannon entropy ${\cal H}$, which
is defined by
\begin{eqnarray*}
{\cal H}_n & := & {\cal H}(R_1,\dots,R_n) \\
 & = & \sum P(r_1,\dots,r_n) \log_2 P(r_1,\dots,r_n), \end{eqnarray*}
and the sum is over all the possible values $r_1,\dots,r_n$ that
$R_1,\dots,R_n$ can take, and by convention
$0 \log_2 0 := \lim_{x \rightarrow 0} x \log_2 x = 0$.

Applying the inequality (\ref{aic inequality}) and the approximation
result (\ref{aic approx}) we find that the average rate
of information production satisfies the inequality
\begin{eqnarray} \label{rate approx}
\overline {\cal R} \geq \limsup_{n \rightarrow \infty}
	\frac{{\cal H}(R_1,\dots,R_n)}{n} =: {\cal R}_S, \end{eqnarray}
where ${\cal R}_S$ is the asymptotic rate at which Shannon
information is produced. ${\cal R}_S$ is another measure of
chaos for the system, and can be interpreted in the following way.
Suppose an experimentalist repeats
the experiment a large number of times, obtaining on each run
a measurement record $R_1,\dots,R_n$. Then the Shannon noiseless
coding theorem \cite{Ash65a} tells us that ${\cal H}(R_1,\dots,R_n)$ is
the smallest average codeword length that can be used to encode the
measurement record. Thus ${\cal R}_S$ represents the asymptotic rate of
growth of the average codeword length needed to encode the measurement record.
See \cite{Schack94b} for a precise account of the distinction
between Shannon entropy and algorithmic information.

In practice, ${\cal R}_S$ was found to be considerably easier to
compute than ${\cal R}$. Furthermore, the inequality (\ref{rate approx})
allows us to obtain bounds on $\overline{\cal R}$ by computing ${\cal R}_S$.
We will now numerically estimate ${\cal R}_S$ for the quantum kicked top.

The {\em kicked top} (see \cite{Haake91a} and references therein) is a
simple system whose classical analogue is known to exhibit chaos.
In units where $\hbar = 1$, the evolution of the quantum kicked top 
from kick to kick is given by the unitary operator
\begin{eqnarray}
U = \exp\left( \! -i \frac{3}{2j}J_z^2\right)
 \exp\left( \! -i\frac{\pi}{2} J_y \right), \end{eqnarray}
where $j$ is the angular momentum quantum number for the system.
In this Letter we have used $j = 18$.

We suppose a projective measurement is performed on the system
immediately after each kick. The projectors used are
\begin{eqnarray} 
P_+ & = & \sum_{m \geq 0} |j,m\rangle \langle j,m| \\
P_- & = & \sum_{m < 0} |j,m\rangle \langle j,m|, \end{eqnarray}
where $J_z|j,m\rangle = m|j,m\rangle$.
This measurement determines whether the $J_z$ component of angular momentum
is non-negative or negative. If the state just before the measurement
is $|\psi\rangle$, then the respective probabilities for the outcomes are
given by the projection postulate as $\langle \psi | P_+ | \psi \rangle$
and $\langle \psi|P_-|\psi \rangle.$

In the numerical examples studied in this Letter the initial states of the
system were chosen to be spin coherent states, which are defined by
\cite{Haake91a}
\begin{eqnarray}
|j,\theta,\phi\rangle = \exp(i\theta(J_x \cos \phi-J_y \sin \phi)) |j,j\rangle.
\end{eqnarray}
In particular, we consider the state
\begin{eqnarray}
|R\rangle = |j=18,\theta=2.25,\phi=0.63\rangle,\end{eqnarray}
for which the means $\langle J_x \rangle, \langle J_y \rangle$ and
$\langle J_z \rangle$ are located in the region of the classical
configuration space for which the classical kicked top is found
to be regular, and the state
\begin{eqnarray}
|C\rangle = |j=18,\theta=1.64,\phi=1.50\rangle,\end{eqnarray}
for which the means $\langle J_x \rangle, \langle J_y \rangle$ and
$\langle J_z \rangle$ are located in the region of the classical
configuration space for which the classical kicked top is found to be
chaotic.

In order to numerically estimate the rate of growth ${\cal R}_S$ of
the Shannon entropy for an initial state $|\psi\rangle$ the following
procedure was used. Suppose $N$ kicks are performed on the system,
and $N$ measurements, one immediately after each kick. This will
result in a sequence of measurement results $Z_1,\dots,Z_N$.
The probability of this measurement history is given by the
projection postulate,
\begin{eqnarray} \begin{array}{lll}
P(Z_1,\dots,Z_N) & = & \langle \psi| U^{\dagger} P_{Z_1} U^{\dagger}
 P_{Z_2} U^{\dagger} \dots U^{\dagger} P_{Z_N} \\
 & & U P_{Z_{N-1}} U \dots P_{Z_1} U|\psi\rangle. \end{array} \end{eqnarray}
Using these probabilities the Shannon information of the measurement record,
\begin{eqnarray}
{\cal H}_n := {\cal H}(Z_1,\dots,Z_n), \end{eqnarray}
was computed for the range $1 \leq n \leq N$.

The results of this procedure when $N = 15$
are shown in figure 1. The lower line plots
${\cal H}_n$ for an initial state $|R\rangle$, and the upper line plots
${\cal H}_n$ for an initial state $|C\rangle$. There are two important things
to note about this graph. First, for both initial states the growth in
$I_n$ settles down to be roughly linear in $n$ very quickly. This has also
been found to be the case with other initial states. Second, the
system started in the state $|R\rangle$ corresponding to the
classically regular region shows a much lower rate of information
production than the state $|C\rangle$, which corresponds to the classically
chaotic region.

The nature of chaos in this system is strikingly illustrated by
choosing $500$ points at random from the $X > 0, Y > 0, Z < 0$
octant of the unit sphere, and calculating
\begin{eqnarray}
\widetilde{{\cal R}}_S :=  \frac{{\cal H}_{15}}{15},\end{eqnarray}
which, assuming ${\cal H}_n$ settles down quickly to a linear
rate of growth, as found above, can be regarded as an approximation to
${\cal R}_S$. This quantity was computed
for each of the initial spin coherent states whose means correspond
to the random points on the sphere. Figure 2 plots
$\widetilde{{\cal R}}_S$ as a function of the angle on the unit sphere
between the point on the unit sphere corresponding to the initial coherent
state, and the point at $\theta = 2.25, \phi = 0.63$, which is an elliptic
fixed point for the classical map, and located deep in the regular
region. The graph shows that the rate of information
production generally increases as this distance increases. That is, the
quantum system becomes more chaotic as the means are moved away from
the elliptic fixed point of the classical map. This is broadly
similar to the classical situation, where the elliptic fixed point is
surrounded by a regular region, which is surrounded by a sea of chaos.
What is striking is that in the quantum case there is a much smoother
transition between the two regimes. Instead of a sharp chaotic and
a sharp regular region there is a continuous transition between
chaotic and regular behavior.

Furthermore, although ${\cal R}$ has not been computed, using the
inequality (\ref{rate approx}) we can place an approximate lower bound on
$\overline{{\cal R}}$, corresponding to the numerically
computed value of $\widetilde{{\cal R}}_S$. We see that
$\overline{{\cal R}}$ is always positive for the initial states plotted in
figure 2.

Two measures of quantum chaos have been developed in this Letter, and
numerical evidence for the existence of a smooth transition between
chaotic and regular behaviour in the quantum kicked top presented. The chief
advantage of using measures determined wholly by the
measurement record is that they can be directly related to the data
available in an experiment. Technically, the approach to quantum chaos
sketched in this Letter may appear similar to that of Schack, Caves
and co-workers \cite{Caves94a,Schack93a,Schack94a,Schack95a,Schack95b},
in the sense that algorithmic information theory is used in both approaches.
However, physically the two approaches are quite
different. The present approach uses measures determined by the
measurement record, whereas Schack, Caves and co-workers focus on measures
determined by the state vector or density operator of the quantum system.

The reader may object that in the present approach to quantum chaos
an appropriate choice of observation scheme may
induce or suppress chaos in the measurement record of systems whose
classical analogues are, respectively, chaotic or not chaotic. For example,
the quantum Zeno effect (see \cite{Peres93a} for an overview and
references) predicts the complete suppression of quantum dynamics when
appropriate measurements are performed on a quantum system. Such
a system will never show chaos in the measurement record, according to our
definitions, even if it is classically chaotic. Other examples may also be
found \cite{Nielsen95f}. A brief answer to such
objections is that the degree of chaos depends on all three of the
initial state, the free dynamics, and the observation scheme.
To obtain classical results we must assume the system is being observed in
an appropriate ``classical'' fashion. This is certainly not the case
for the quantum Zeno effect. However, this should not be taken to imply
that observation is the sole cause of chaos in the measurement record. As
we have seen, chaos in the measurement record depends also on the system's
initial condition, and it can be shown \cite{Nielsen95f} that it depends
on the free dynamics as well. These issues will be dealt with more
completely in \cite{Nielsen95f}.

The approach to quantum chaos sketched in this Letter may be extended in
a number of ways. More general measurement schemes may be considered,
using formalisms such as quantum trajectories
\cite{Gardiner92a,Molmer93a,Carmichael93a,Nielsen95a}. In particular,
it is necessary to consider imperfect measurements, which introduce
extra noise into the measurement record. Closely related is the idea
of studying the effect of interactions with the environment other
than the coupling to a measuring device. Finally, it is interesting to
examine more thoroughly how varying the initial state, the
system Hamiltonian and the observation scheme affect the degree
of chaos seen in the measurement record. These and other issues
will be dealt with in a more extensive paper \cite{Nielsen95f}.

\acknowledgements

I would like to thank J. K. Breslin, J. Twamley and G. J. Milburn for
many useful discussions about this work. Thanks also to
C. M. Caves and Matthew R. Semak who read the manuscript and provided
helpful comments. This work was supported by an Australian
Postgraduate Award and a Fulbright Scholarship.


\figure{FIG 1. Shannon information as a function of time for the kicked
top started in the states $|R \rangle$ (bottom) and $|C \rangle$ (top).}

\figure{FIG 2. Rate of production of Shannon information as a function of
distance from the elliptic fixed point in the $X >0, Y>0, Z<0$ octant of the
unit sphere.}

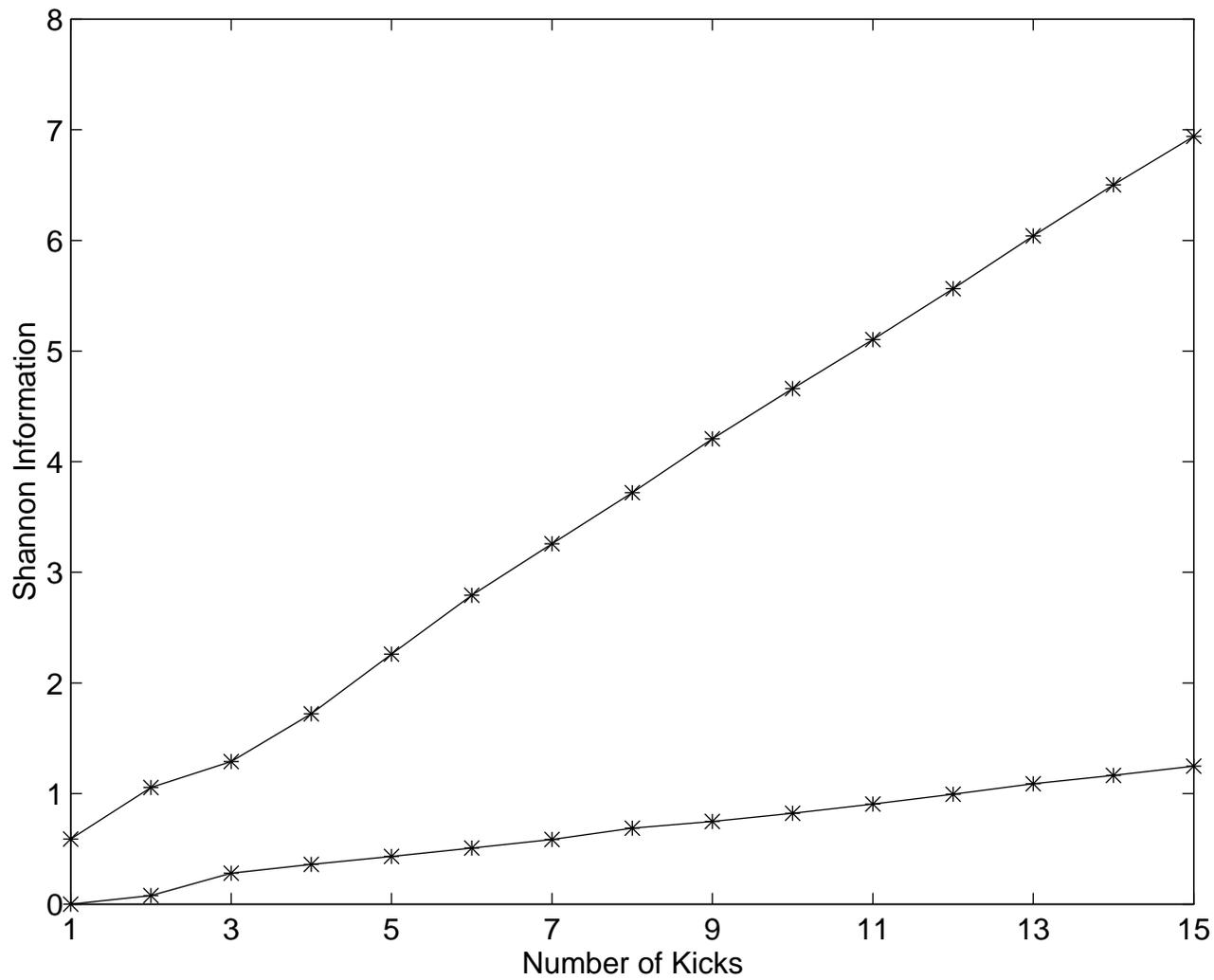
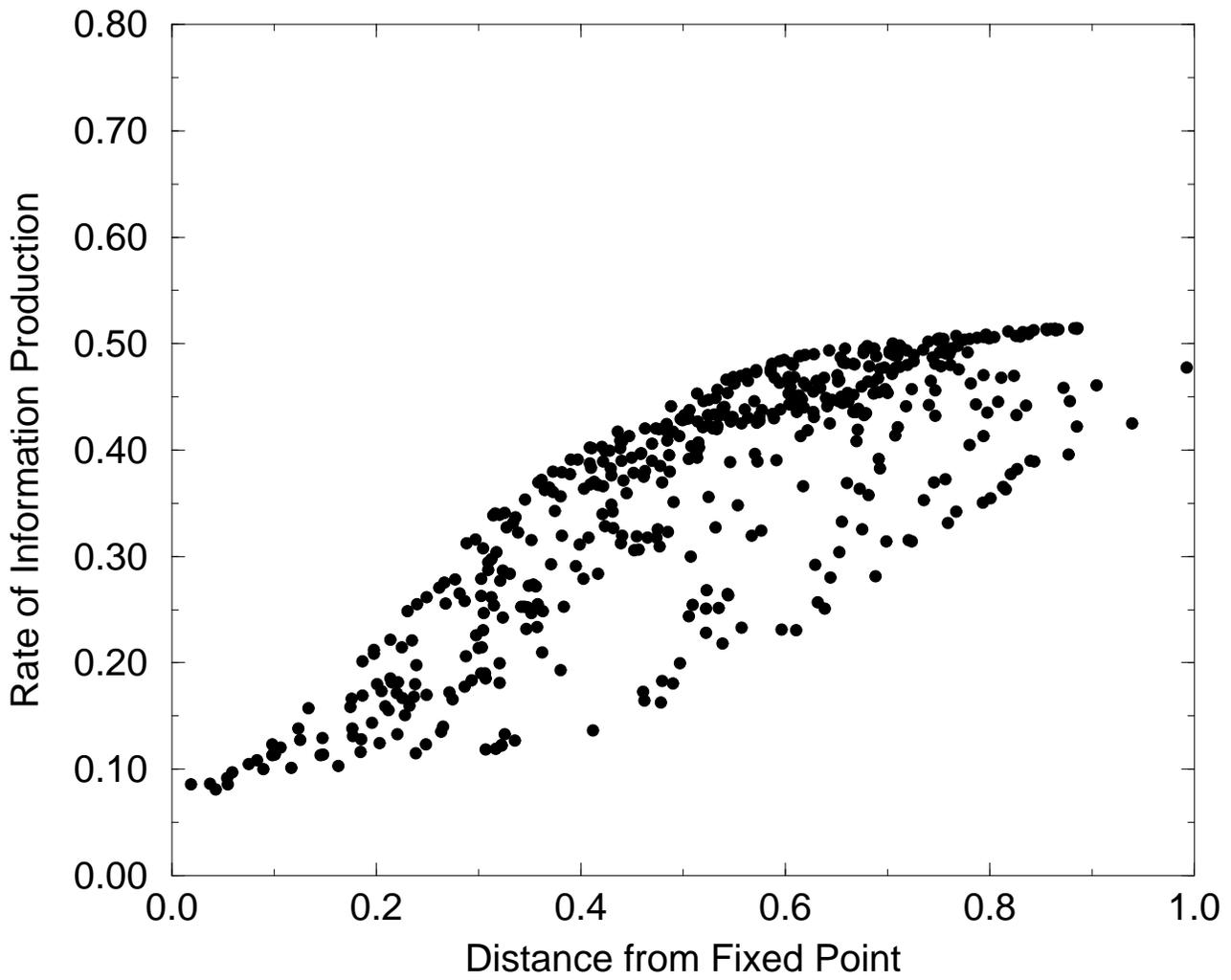
\end{document}